\documentclass[12pt]{article}
\usepackage{amsmath,amssymb}

\newcommand{\bea}{\begin{eqnarray}}
\newcommand{\eea}{\end{eqnarray}}
\newcommand{\be}{\begin{equation}}
\newcommand{\ee}{\end{equation}}
\newcommand{\vs}[1]{\vspace{#1 mm}}

\newcommand{\dsl}{\pa \kern-0.5em /}

\newcommand{\pa}{\partial}

\newcommand{\nn}{\nonumber\\}

\begin{document}
\topmargin 0mm
\oddsidemargin 0mm

\begin{flushright}

USTC-ICTS/PCFT-24-59\\

\end{flushright}

\vspace{2mm}

\begin{center}

{\Large \bf The compactified D-brane cylinder amplitude \\ and T duality\\}

\vs{10}

{\large J. X. Lu}

\vspace{4mm}

{\em
Interdisciplinary Center for Theoretical Study\\
 University of Science and Technology of China, Hefei, Anhui
 230026, China\\
 \medskip
 Peng Huanwu Center for Fundamental Theory, Hefei, Anhui 230026, China\\ 
 %\vs{4}
}

\end{center}

\vs{10}

\begin{abstract}
In this paper, we  address how to implement T duality to the  closed string tree cylinder amplitude between a Dp brane and a Dp$'$ brane with $p - p' = 2 \,n$. For this, we first compute the closed string tree cylinder amplitude between these two D branes with common longitudinal and transverse circle compactifications. We then show explicitly how to perform a T duality  to this amplitude along  either   a longitudinal or a transverse compactified direction to both branes.  In the decompactification limit, we show that either the compactified cylinder amplitude or the T dual compactified cylinder one gives the known non-compactified one as expected.
 \end{abstract}

\newpage

% Use the \preprint command to place your local institutional report
% number in the upper righthand corner of the title page in preprint mode.
% Multiple \preprint commands are allowed.
% Use the 'preprintnumbers' class option to override journal defaults
% to display numbers if necessary
%\preprint{}

% insert suggested PACS numbers in braces on next line
%\pacs{}
% insert suggested keywords - APS authors don't need to do this
%\keywords{}
%\maketitle must follow title, authors, abstract, \pacs, and \keywords
%\maketitle

% body of paper here - Use proper section commands
% References should be done using the \cite, \ref, and \label commands
\section{Introduction}
In general, we know that a Dp brane will become a D(p - 1) brane if a T-duality is performed along its longitudinal direction and a D(p + 1) brane if the T duality is done along a direction transverse to it.  However, it is not  clear, to the best of my knowledge, how to implement this T-duality to a closed string  tree cylinder amplitude between a Dp and a Dp$'$, placed parallel at a separation, with $p - p' = 2 n$ and $2n \le p \le 8$ (without loss of generality, we assume from now on $p > p'$)\footnote{It is known that when $p - p' = 2 n$ and $n = 0, 2$ this amplitude vanishes due to the underlying systems preserving 1/2 and 1/4 spacetime supersymmetry, respectively.  For $n = 4$, we have only one system consisting of a D8 and a D0 and the corresponding amplitude vanishes also. However, it is known that this system has various subtle issues \cite{Danielsson:1997wq, Bergman:1997gf, Billo:1998np, Lu:2009gm} and we will not consider it in this paper for simplicity. So in this paper, we limit ourselves to $n = 0, 1, 2, 3$. }.

We try to address this in the present paper.  As will become clear, we can perform a T duality along a direction transverse to both branes for such systems if we periodically place them along this direction to give an effective circle compactification with a  radius $R \to 0$ and then perform the T duality to give a system of D(p + 1) and D(p$'$ + 1) without compactification.  We will show that this T duality works only in the $R \to 0$ limit. However, this same process will not work if a T duality is performed along a longitudinal direction common to both branes. We will explain the reason behind.

So in order to properly perform a T duality on the amplitude in general, we need to compute the corresponding amplitude with at least one compactified direction which is either longitudinal or transverse to both branes and the T duality is performed along this direction.

For this, we compute in this paper the general  closed string tree cylinder amplitude between a Dp and a Dp$'$, as specified above,  with $k \le p'$ spatial longitudinal circle compactifications of radii $r_{1}, r_{2}, \cdots r_{k}$ and $l \le 9 - p$ transverse circle compactifications of  radii $R_{1}, R_{2}, \cdots R_{l}$.  This amplitude will serve as a starting point  for us to perform a T duality along either a longitudinal or a transverse direction. 

This paper is arranged as follows.  In the following section, we will give a brief review of Dp brane bound state representation as given, for example, in the nice review  \cite{DiVecchia:1999mal}. We will use this representation to demonstrate in this section how each sector of the matter part in the D-brane boundary state transforms under a T duality along either a longitudinal or transverse direction to the brane. This shows that the only part for which we need to take into consideration when one such direction is compactified  is the bosonic matter zero-mode to the boundary state.  The corresponding compactified bosonic matter zero mode boundary state is already given in \cite{DiVecchia:1999fje} when the compactified circles have the same radius. We generalize this boundary state to one with $k \le p - 2 n$ spatial longitudinal compactifications with the respective radii $r_{1}, \cdots r_{k}$ and $l \le 9 -p$ transverse compactifications with the respective radii $R_{1}, \cdots R_{l}$ as specified earlier.  With such compactified boundary states, we compute the corresponding  closed string cylinder amplitude between a Dp and a Dp$'$ with $p - p' = 2 n$. In section 4, we address how to implement a T duality along either the longitudinal or the transverse direction to the cylinder amplitude computed in the previous section and check the consistency along with other discussions. We conclude this paper in section 5.

\section{The D-brane boundary state and its T duality}

D-branes are a sort of non-perturbative with respect to the fundamental string or F-string.  We specify our discussion here in Type II superstring theories in 10 dimensions. For a weak string coupling, D-branes can be viewed as rigid at least to the scale up to stringy one. In this case, a Dp brane can be represented as a coherent state of closed string excitations, called Dp brane boundary state. This has been discussed in great detail in the nice review \cite{DiVecchia:1999mal}. We simply quote here the boundary state which can be expressed as the product of a matter part and a ghost part
\be\label{Dp-bs}
  |B, \eta\rangle = |B_{\rm mat}, \eta\rangle |B_{\rm g}, \eta\rangle,
 \ee
 where 
 \be\label{mgbs}
 |B_{\rm mat}, \eta\rangle = |B_X \rangle|B_\psi, \eta\rangle,  \quad |B_{\rm g},\eta\rangle = |B_{\rm gh}\rangle |B_{\rm sgh}, \eta\rangle,
 \ee
with here $\eta = \pm$.     In the above, 
\be\label{xbs}
 |B_X\rangle ={\rm exp} (-\sum_{n =1}^\infty \frac{1}{n} \alpha_{- n} \cdot S
\cdot {\tilde \alpha}_{ - n}) |B_X\rangle_0,
\ee
 and 
 \be\label{bspsinsns}
 |B_\psi, \eta\rangle_{\rm NS} = - {\rm i}~ {\rm exp} (i \eta
\sum_{m = 1/2}^\infty \psi_{- m} \cdot S \cdot {\tilde \psi}_{- m})|0\rangle,
\ee for the NS-NS sector and 
\be\label{bspsirr}
 |B_\psi,\eta\rangle_{\rm R} = - {\rm exp} (i \eta \sum_{m = 1}^\infty
\psi_{- m} \cdot S \cdot {\tilde \psi}_{- m}) |B,
\eta\rangle_{0\rm R},
\ee for the R-R sector. The ghost boundary states are the standard ones as given in \cite{Billo:1998np} and they are irrelevant to T dualities we are interested in this paper. For this reason, we will not list them here. The S-matrix in the absence of D brane worldvolume fluxes  is simply 
\be\label{S-matrix}
S_{p} = \left(\eta_{\alpha\beta},  - \delta_{ij}\right)
\ee
 and the zero-mode boundary states are, respectively, 
\be\label{bzm}
|B_X\rangle_0 =  \frac{c_p}{2}\, \delta^{(9 - p)} (q^i - y^i) \prod_{\mu = 0}^{9} |k^\mu = 0\rangle,
\ee 
for the bosonic sector with the overall normalization $c_p = \sqrt{\pi}\left(2\pi \sqrt{\alpha'}\right)^{3 - p}$, and 
\be\label{rrzm}
|B_\psi, \eta\rangle_{0\rm R} = \left(C \Gamma^0 \Gamma^1
\cdots \Gamma^p \frac{1 + {\rm i} \eta \Gamma_{11}}{1 + {\rm i} \eta
} \right)_{AB} |A\rangle |\tilde B\rangle,
\ee
 for the R-R sector.   
 
  In the above, the Greek indices $\alpha, \beta, \cdots$ label the
world-volume directions $0, 1, \cdots, p$ along which the Dp
brane extends, while the Latin indices $i, j, \cdots$ label the
directions transverse to the brane, i.e., $p + 1, \cdots, 9$. 
 We also have denoted by $y^i$ the
positions of the D-brane along the transverse directions, by $C$ the
charge conjugation matrix. 
$|A \rangle |\tilde B\rangle$ stands for the spinor vacuum of the R-R
sector. 
Note that the $\eta$ in the above
denotes either sign $\pm$ or the worldvolume Minkowski flat metric and should be clear from the content.

 The above boundary states $|B, \eta\rangle$ with $\eta = \pm$ are not the ones to give the Dp brane description but the following combinations of the so-called Gliozzi-Scherk-Olive (GSO) projected ones 
\bea\label{gsop-Dp-bs}
 &&|B \rangle_{\rm NSNS} = \frac{1}{2} \left[|B, + \rangle_{\rm NSNS} - |B, - \rangle_{\rm NSNS}\right],\nn
 &&|B \rangle_{\rm RR} = \frac{1}{2} \left[|B, + \rangle_{\rm RR} + |B, - \rangle_{\rm RR}\right],
 \eea 
in the respective so-called NS-NS sector and R-R sector. These are the boundary states we are going to use in computing the closed string cylinder amplitude between a Dp brane and a Dp$'$ brane with $p - p' = 2 n$ as specified earlier. 

Our purpose is to seek how to perform a T duality to the amplitude. As mentioned earlier,  we need to consider the corresponding compactified D brane boundary states. In the following, we will show that the only thing indeed relevant is the bosonic matter zero-mode boundary state.  For this,  we pose here to recall briefly the simplest T duality acting on a  closed string in Type II  superstring theories.

For this, we compactify a spatial direction of the 10 dimensional spacetime, say $X$, to give a circle of a radius $R$ in one theory (say IIA) with its closed string wrapping this circle $W$ times and having a quantized momentum $K$, and to give the other circle of $\tilde R$ in the other theory (say IIB)  with its closed string having the winding $\tilde W$ and momentum $\tilde K$. If the radius $R$ is related to the other radius $\tilde R$ via $R \tilde R = \alpha'$ and the winding and momentum modes are exchanged via $K \leftrightarrow \tilde W$ and $W \leftrightarrow \tilde K$, we  have then  the equivalent relation ${\rm IIA} \leftrightarrow {\rm IIB}$ via this T duality.   The above simple picture of T duality can be generalized to all modes for both bosonic and fermionic modes in the following way
\be\label{T-dual-rm}
\left\{\begin{array}{l}
X_{R} (\tau -\sigma) \to - X_{R} (\tau - \sigma); \quad \psi_{R} (\tau - \sigma) \to - \psi_{R} (\tau - \sigma), \\

(\alpha_{n} \to - \alpha_{n}, x \leftrightarrow  \hat x; \quad  \psi_{t} \to - \psi_{t}) \end{array}\right. 
\ee
for the right mover, while for the left mover 
\be\label{T-dual-lm}
\left\{\begin{array}{l}
X_{L} (\tau + \sigma)  \to X_{L} (\tau + \sigma); \quad \psi_{L} (\tau + \sigma) \to \psi_{L} (\tau + \sigma),  \\
(\tilde \alpha_{n} \to \tilde\alpha_{n},\, x \leftrightarrow \hat x; \quad \tilde\psi_{t} \to \tilde\psi_{t}). \end{array} \right.
\ee
where the index $n$ is an integer while the index $t$ is an integer in R sector and a half-integer in NS sector.   Note that 
\be
\alpha_{0} = \sqrt{\frac{\alpha'}{2}} \left(\frac{K}{R} - \frac{W R}{\alpha'}\right), \quad \tilde\alpha_{0}  = \sqrt{\frac{\alpha'}{2}} \left(\frac{K}{R} + \frac{W R}{\alpha'}\right).
\ee
Under T duality,  the generators of Virasoro algebra of the matter part  for either left or right mover remain invariant. We can see this easily from the following zero modes as examples
\bea\label{Virasoro-g-zero}
L^{\rm matter}_{0} &=& \frac{\alpha'}{4} \left(\frac{K}{R} - \frac{W R}{\alpha'}\right)^{2} + \frac{\alpha'}{4} \hat p^{2} + \sum_{n = 1}^{\infty} \alpha_{- n} \cdot \alpha_{n} + \sum_{t > 0}^{\infty} \psi_{- t} \cdot \psi_{t},\nn
\tilde L^{\rm matter}_{0} &=& \frac{\alpha'}{4} \left(\frac{K}{R} + \frac{W R}{\alpha'}\right)^{2} + \frac{\alpha'}{4} \hat p^{2} + \sum_{n = 1}^{\infty} \tilde\alpha_{- n} \cdot \tilde\alpha_{n} + \sum_{t > 0}^{\infty} \tilde\psi_{- t} \cdot \tilde\psi_{t},
\eea
where $\hat p$ labels the momentum along the non-compactified directions.  Note  that in the R sector, we have fermionic zero modes $\psi_{0}$ and $\tilde \psi_{0}$ and they transform under the T duality along the compactified direction as
\be
\psi_{0} \to - \psi_{0}, \qquad \tilde \psi_{0} \to \tilde\psi_{0}.
\ee
Note also $\psi_{0} \sim \Gamma$ with $\Gamma$ the Dirac matrix along the compactified direction\footnote{\label{fn1}We adopt here that the 10 dimensional Dirac matrices are all real with the spatial ones symmetric and the temporal one antisymmetric. Their explicit representations are given, for example, in the Appendix in \cite{DiVecchia:1999fje}. The charge conjugate $C$ is also given explicitly there  which remains invariant under T duality.}. This implies that $\Gamma \to - \Gamma$ under the T duality. So this further implies that the chiral operator $\Gamma_{11} \to - \Gamma_{11}$.  So under the T duality, the chirality for one of the two chiral spinors in Type II flips which gives 
\be
{\rm IIA} \leftrightarrow {\rm IIB}.
\ee

With this brief introduction of T duality, we now examine how the boundary state given above transforms under it.  For this, let us first examine how each of the exponential factors due to either the bosonic or the fermionic oscillators in the respective boundary states of matter part transforms under the T -duality. Given (\ref{T-dual-rm}) and (\ref{T-dual-lm}), it is easy to see that the change of each of these factors under the T duality  amounts to the change of the $S_{p}$ matrix (\ref{S-matrix}) as follows
\be
S_{p} = \left(\eta_{\alpha\beta}, - \delta_{ij}\right) \to S_{p - 1} = \left(\eta_{\alpha' \beta'}, - \delta_{i' j'}\right),
\ee
where without loss of generality (for simplicity and for convenience) we have chosen the compactification $X = X^{p}$, i.e., along the brane direction, with $\alpha', \beta' = 0, 1, \cdots p - 1$ and $i', j' = p, p + 1, \cdots 9$ or 
\be
S_{p} = \left(\eta_{\alpha\beta}, - \delta_{ij}\right) \to S_{p + 1} =  \left(\eta_{\alpha' \beta'}, - \delta_{i' j'}\right),\
\ee
where we have chosen the compactification $X = X^{p + 1}$, i.e., along the direction transverse to the brane, with now $\alpha', \beta' = 0, 1, \cdots, p + 1$ and $i', j' = p + 2, p + 3, \cdots 9$.  Either of these is consistent with T duality.  We come now to examine how the fermionic zero mode boundary state transforms under T duality. If we choose $X = X^{p}$ for simplicity, we have, under a T duality along this direction, $\Gamma^{p} \to - \Gamma^{p}$ and $\Gamma_{11} \to - \Gamma_{11}$.  We  have  then from (\ref{rrzm})
\bea\label{T-dual-rrzm}
|B^{p}_\psi, \eta\rangle_{0\rm R} &=& \left(C \Gamma^0 \Gamma^1
\cdots \Gamma^p \frac{1 + {\rm i} \eta \Gamma_{11}}{1 + {\rm i} \eta
} \right)_{AB} |A\rangle |\tilde B\rangle\nn
 &\to&  -  \left(C \Gamma^0 \Gamma^1
\cdots \Gamma^p \frac{1 - {\rm i} \eta \Gamma_{11}}{1 + {\rm i} \eta
} \right)_{AB} |A\rangle |\tilde B\rangle \nn
&=& -  \left(C \Gamma^0 \Gamma^1
\cdots \Gamma^{p -1} \frac{1 + {\rm i} \eta \Gamma_{11}}{1 + {\rm i} \eta
} \right)_{AD} |A\rangle \left(\Gamma^{p}\right)_{DB} |\tilde B\rangle,
\eea
where we have used  $C \to C$ under the T duality as mentioned in footnote (\ref{fn1}).    Note also that we can choose $(\Gamma_{11})_{AB} |B\rangle = |A \rangle$ and $(\Gamma_{11})_{AB} |\tilde B\rangle = \pm |\tilde A\rangle$ where the sign depends on which theory (IIA or IIB) we start with. For $` - '$ sign, we start with IIA while for $`+'$ sign we start with IIB. Denoting $|{\tilde {\cal B}}\rangle = - \left(\Gamma^{p}\right)_{BD} |\tilde D \rangle$, we have $\left(\Gamma_{11}\right)_{AB} |{\tilde {\cal B}}\rangle = \mp |{\tilde {\cal A}}\rangle$. In other words, after the T duality, we transform $|\tilde B\rangle$ to $|{\tilde {\cal B}}\rangle$ with an opposite chirality\footnote{Given that $\Gamma_{11}$ is a symmetric matrix, we can also apply this to $|A\rangle$.}, i.e. ${\rm IIA \leftrightarrow IIB}$, as expected. Given all this, we have from (\ref{T-dual-rrzm}) under the T duality
\bea
|B^{p}_\psi, \eta\rangle_{0\rm R} &\to& -  \left(C \Gamma^0 \Gamma^1
\cdots \Gamma^{p -1} \frac{1 + {\rm i} \eta \Gamma_{11}}{1 + {\rm i} \eta
} \right)_{AD} |A\rangle \left(\Gamma^{p}\right)_{DB} |\tilde B\rangle\nn
&=&  \left(C \Gamma^0 \Gamma^1
\cdots \Gamma^{p -1} \frac{1 + {\rm i} \eta \Gamma_{11}}{1 + {\rm i} \eta
} \right)_{AB} |A\rangle | {\tilde {\cal B}}\rangle\nn
&=&   |B^{p - 1}_\psi, \eta\rangle_{0\rm R}.
\eea
So under the T duality along a longitudinal direction of the Dp brane, we have $|B^{p}_\psi, \eta\rangle_{0\rm R} \to   |B^{p - 1}_\psi, \eta\rangle_{0\rm R}$ (${\rm IIA \leftrightarrow IIB}$) and the R sector zero mode boundary state transforms as expected.  If we choose instead to perform a T duality along a direction, say  $X = X^{p + 1}$, transverse to the p-brane, we still have $\Gamma_{11} \to - \Gamma_{11}$ due to $\Gamma^{p + 1} \to - \Gamma^{p + 1}$.  Since $(\Gamma^{p + 1})^{2} = \mathbf{I}_{32\times 32}$ with $\mathbf{I}_{N \times N}$ the $N \times  N$  unit matrix, we then have, from (\ref{rrzm})
\bea\label{T-dual-rrzmt}
|B^{p}_\psi, \eta\rangle_{0\rm R} &=& \left(C \Gamma^0 \Gamma^1
\cdots \Gamma^p \frac{1 + {\rm i} \eta \Gamma_{11}}{1 + {\rm i} \eta
} \right)_{AB} |A\rangle |\tilde B\rangle\nn
&\to&  \left(C \Gamma^0 \Gamma^1
\cdots \Gamma^p \frac{1 - {\rm i} \eta \Gamma_{11}}{1 + {\rm i} \eta
} \right)_{AB} |A\rangle |\tilde B\rangle\nn
&=&  \left(C \Gamma^0 \Gamma^1
\cdots \Gamma^p \frac{1 - {\rm i} \eta \Gamma_{11}}{1 + {\rm i} \eta
} \left(\Gamma^{p + 1}\right)^{2} \right)_{AB} |A\rangle |\tilde B\rangle\nn
&=&    \left(C \Gamma^0 \Gamma^1
\cdots \Gamma^p  \Gamma^{p + 1}\frac{1 + {\rm i} \eta \Gamma_{11}}{1 +  {\rm i} \eta
} \right)_{AD} |A\rangle \left(\Gamma^{p + 1}\right)_{DB}|\tilde B\rangle\nn
&=&    \left(C \Gamma^0 \Gamma^1
\cdots \Gamma^p  \Gamma^{p + 1}\frac{1 + {\rm i} \eta \Gamma_{11}}{1 +  {\rm i} \eta
} \right)_{AB} |A\rangle |{\tilde {\cal B}}\rangle\nn
&=&  |B^{p + 1}_\psi, \eta\rangle_{0\rm R},
\eea
where $|{\tilde {\cal B}}\rangle \equiv \left(\Gamma^{p + 1}\right)_{BD}|\tilde D\rangle$.  So under the T duality along a direction transverse to the Dp brane,  by a similar token, we have $|B^{p}_\psi, \eta\rangle_{0\rm R} \to  |B^{p + 1}_\psi, \eta\rangle_{0\rm R}$ and ${\rm IIA \leftrightarrow IIB}$, also as expected. So the only thing left is how the bosonic zero mode boundary state transforms under the T duality either along a longitudinal or transverse direction. 

Given that the bosonic zero modes change indeed from the non-compactified case to the compactified one, as given earlier, we expect that the bosonic zero mode boundary state changes also as already stressed and considered in \cite{DiVecchia:1999fje}.  We list the bosonic zero-mode boundary state for a Dp brane with $k$ longitudinal compactified directions 
of radii $r_{\alpha_{i}}$ ($i = 1, \cdots k$) and $l$ transverse compactified directions of radii $R_{j_{m}}$ ($m = 1, \cdots, l$), generalized to different radii from that given in \cite{DiVecchia:1999fje}, as
\bea\label{compactified-pbs}
|\Omega_{p}\rangle &=& {\cal N}_{p} \prod_{i = 1}^{k}\left[ \sum_{\omega_{\alpha_{i}}} e^{- i\, y^{\alpha_{i}} \omega_{\alpha_{i}} r_{\alpha_{i}}/\alpha' } |n_{\alpha_{i}} = 0, \omega_{\alpha_{i}} \rangle\right] |k^{0} = 0, \hat k^{\parallel} = 0\rangle\nn
&\,& \times  \prod_{m = 1}^{l} \left[\sum_{n_{j_{m}}} e^{- i\, y^{j_{m}} n_{j_{m}} /R_{j_{m}}} |n_{j_{m}}, \omega_{j_{m}} = 0 \rangle \right] \hat \delta^{(\perp)} \left( \hat q^{\perp} - \hat y^{\perp}\right) |\hat k^{\perp} = 0\rangle,
\eea
where we use `$\,\hat\, $' to denote those directions, either longitudinal or transverse, which are not compactified.  In the above, the normalization factor
\be\label{normalization}
{\cal N}_{p} =  \frac{c_p}{2} \prod_{i = 1}^{k} \left(\frac{2 \pi r_{\alpha_{i}}}{\Phi_{\alpha_{i}}}\right)^{1/2} \prod_{m = 1}^{l} \left(\frac{1}{2 \pi R_{j_{m}}\Phi_{j_{m}}}\right)^{1/2}.
\ee 
Following \cite{DiVecchia:1999fje}, we introduce, for convenience, the `position' and `momentum' operators, respectively, for the  momentum and winding degrees of freedom,  as
\be 
\left[q^{\mu}_{\omega}, p^{\nu}_{\omega}\right] = i \delta^{\mu\nu}, \quad \left[q^{\mu}_{n}, p^{\nu}_{n}\right] = i \delta^{\mu\nu},
\ee 
where $\mu, \nu$ are along either the longitudinal or the transverse compactified spatial directions.  We then have
\be\label{momentum-eigen}
p^{\alpha}_{n} |\Omega_{p}\rangle = 0, \qquad p^{j}_{\omega} |\Omega_{p} \rangle = 0,
\ee
where $\alpha$ represents one of $\alpha_{i}$ and $j$ represents one of $j_{m}$. By denoting the eigenstate $|n_{\nu}, \omega_{\nu}\rangle$ of the respective `momentum' operators, we have
\be
p^{\mu}_{n} |n_{\mu}, \omega_{\mu} \rangle = \frac{n_{\mu}}{a_{\mu}}  |n_{\mu}, \omega_{\mu} \rangle, \quad p^{\mu}_{\omega} |n_{\mu}, \omega_{\mu} \rangle = \frac{\omega_{\mu}a_{\mu}}{\alpha'} |n_{\mu}, \omega_{\mu} \rangle,
\ee
where $a_{\mu}$ is the radius of the compactified direction which can be either one of $r_{\alpha_{i}}$ or one of $R_{j_{m}}$ mentioned earlier.  Given the above, it is easy to write down the state as
\be
|n_{\mu}, \omega_{\mu}\rangle \equiv e^{i q^{\mu}_{n} n_{\mu}/a_{\mu}} \, e^{i q^{\mu}_{\omega} \omega_{\mu} a_{\mu} /\alpha'} |0, 0\rangle,
\ee
with $|0, 0\rangle$ denoting the zero-momentum and zero-winding state. The normalization of this state is given 
\be
\langle n'_{\mu}, \omega'_{\mu} |n_{\mu}, \omega_{\mu} \rangle = \Phi_{\mu}\, \delta_{n'_{\mu}, n_{\mu}}\, \delta_{\omega'_{\mu}, \omega_{\mu}},
\ee
where $\Phi_{\mu}$ is the so called `self-dual' volume\cite{DiVecchia:1999fje}  which has the following properties 
\be
\Phi_{\mu} = 2 \pi a_{\mu}\,\, {\rm if}\,\,  a_{\mu} \to \infty; \quad \Phi_{\mu} = \frac{2\pi \alpha'}{a_{\mu}} \,\, {\rm if}\,\,  a_{\mu} \to 0.
\ee
Note also that in the decompactification limit we have when $r_{\alpha} \to \infty$ 
\be\label{winding}
\sum_{\omega_{\alpha}} e^{i \left(q_{\omega}^{\alpha} - y^{\alpha}\right) \omega_{\alpha} r_{\alpha}/\alpha' } |0, 0\rangle \to |0, 0\rangle
\ee
where $\alpha$ is one of $\alpha_{i}$, and when $R_{j} \to \infty$
\bea\label{momentum}
&&\sum_{n_{j}} e^{i \left(q^{j}_{n} - y^{j}\right) n_{j} /R_{j}} |0, 0\rangle \to  R_{j} \int d k^{j} \, e^{i \left(q^{j} - y^{j}\right) k^{j}} |0, 0\rangle =\nn
&& 2\pi R_{j} \int \frac{d k^{j}}{2\pi} e^{i \left(q^{j} - y^{j}\right) k^{j}} |0, 0\rangle = 2 \pi R_{j} \delta \left(q^{j} - y^{j}\right) |0, 0\rangle,
\eea
where $j$ is one of $j_{m}$.  In the above, when taking $r_{\alpha} \to \infty$,  only the $\omega_{\alpha} = 0$ term survives in the sum in (\ref{winding}) while in (\ref{momentum}) we have replaced the sum by an integral of $k$ given by $k^{j} = n_{j} /R_{j}$ when $R_{j} \to \infty$.

With the above preparation, we now come to perform a T duality  along either a longitudinal or a transverse direction to the zero-mode state $|\Omega_{p}\rangle$ to see if it is consistent with our expectation.  Let us begin with a T duality along a longitudinal direction first.  Without loss of generality, let us perform this T-duality along the $\alpha_{k}$-direction. We then need to send $r_{\alpha_{k}} \to \alpha'/r_{\alpha_{k}}$ and $n_{\alpha_{k}} \leftrightarrow \omega_{\alpha_{k}}$. With this, we have
\bea
|\Omega_{p}\rangle &\to& \tilde{\cal N}_{p}  \prod_{i = 1}^{k - 1} \left[\sum_{\omega_{\alpha_{i}}} e^{- i y^{\alpha_{i}} \omega_{\alpha_{i}} r_{\alpha_{i}} /\alpha'} |n_{\alpha_{i}} = 0, \omega_{\alpha_{i}} \rangle\right] |k^{0} = 0, \hat k^{\parallel} = 0\rangle \nn
&\,&  \times  \left(\sum_{n_{\alpha_{k}}} e^{- i y^{\alpha_{k}} n_{\alpha_{k}}/r_{\alpha_{k}}} |n_{\alpha_{k}}, \omega_{\alpha_{k}} = 0 \rangle\right)\nn
&\,&  \times  \prod_{m = 1}^{l} \left[\sum_{n_{j_{m}}} e^{- i\, y^{j_{m}} n_{j_{m}} /R_{j_{m}}} |n_{j_{m}}, \omega_{j_{m}} = 0 \rangle \right] \hat \delta^{(\perp)} \left( \hat q^{\perp} - \hat y^{\perp}\right) |\tilde k^{\perp} = 0\rangle\nn
&=&\tilde{\cal N}_{p}  \prod_{i = 1}^{k - 1}\left[ \sum_{\omega_{\alpha_{i}}} e^{- i\, y^{\alpha_{i}} \omega_{\alpha_{i}} r_{\alpha_{i}}/\alpha' } |n_{\alpha_{i}} = 0, \omega_{\alpha_{i}} \rangle\right] |k^{0} = 0, \hat k^{\parallel} = 0\rangle\nn
&\,& \times  \prod_{m = 1}^{l + 1} \left[\sum_{n_{j_{m}}} e^{- i\, y^{j_{m}} n_{j_{m}} /R_{j_{m}}} |n_{j_{m}}, \omega_{j_{m}} = 0 \rangle \right] \hat \delta^{(\perp)} \left( \hat q^{\perp} - \hat y^{\perp}\right) |\tilde k^{\perp} = 0\rangle,
\eea
where in the last equality we have taken $n_{j_{l + 1}} = n_{\alpha_{k}} $, $j_{l + 1} = \alpha_{k}$ and $R_{j_{l + 1}} =  r_{\alpha_{k}}$.  Note also under this T duality, we have
\bea
{\cal N}_{p} \to \tilde{\cal N}_{p} &=&  \frac{c_{p}}{2} \left(\frac{2 \pi \alpha'}{ r_{\alpha_{k}}\Phi_{\alpha_{k}}}\right)^{1/2}  \prod_{i = 1}^{k - 1} \left(\frac{2 \pi r_{\alpha_{i}}}{\Phi_{\alpha_{i}}}\right)^{1/2} \prod_{m = 1}^{l} \left(\frac{1}{2 \pi R_{j_{m}}\Phi_{j_{m}}}\right)^{1/2}\nn
&=& \frac{c_{p}}{2} \left(2\pi \sqrt{\alpha'}\right)  \left(\frac{1}{2\pi  r_{\alpha_{k}}\Phi_{\alpha_{k}}}\right)^{1/2}   \prod_{i = 1}^{k - 1} \left(\frac{2 \pi r_{\alpha_{i}}}{\Phi_{\alpha_{i}}}\right)^{1/2} \prod_{m = 1}^{l} \left(\frac{1}{2 \pi R_{j_{m}}\Phi_{j_{m}}}\right)^{1/2}\nn
&=& \frac{c_{p - 1}}{2}  \prod_{i = 1}^{k - 1} \left(\frac{2 \pi r_{\alpha_{i}}}{\Phi_{\alpha_{i}}}\right)^{1/2} \prod_{m = 1}^{l + 1} \left(\frac{1}{2 \pi R_{j_{m}}\Phi_{j_{m}}}\right)^{1/2} = {\cal N}_{p - 1}.
\eea
With the above, we have, under this T duality, the expected transformation 
\be
|\Omega\rangle_{p} \to |\Omega\rangle_{p - 1}.
\ee
By a similar token, one can also show, when a T duality is performed along a transverse compactified direction,  
\be
|\Omega\rangle_{p} \to |\Omega\rangle_{p + 1},
\ee
which is also expected. 

\section{The compactified D-brane cylinder amplitude}
With the preparation given in the previous section, we are now ready to compute the closed string tree cylinder amplitude between a Dp and a Dp$'$ with $p - p' = 2 n$. The two D branes are placed parallel at a separation along their common non-compactified transverse directions. Here the compactifications are respect to the longitudinal and transverse directions common to both of the D branes. In other words, for the compactified boundary state given in the previous section, we need to restrict $k \le p'$ and $l \le 9 - p$ as discussed in Introduction. 

The cylinder amplitude between a Dp brane and a Dp$'$ brane, placed parallel at a separation $y$, can be calculated via
 \be\label{ampli}
 \Gamma_{\rm Dp'|Dp} = \langle B^{p'} | D |B^{p} \rangle,
 \ee
 where $D$ is the closed string propagator defined as 
\be\label{prog}
 D =
\frac{\alpha'}{4 \pi} \int_{|z| \le 1} \frac{d^2 z}{|z|^2} z^{L_0}
 {\bar z}^{{\tilde L}_0}.
 \ee 
 Here $L_0$ and ${\tilde L}_0$ are the respective left and right mover total zero-mode Virasoro
generators of matter fields, ghosts and superghosts. For example, $L_0 = L^X_0 + L_0^\psi + L_0^{\rm gh} + L_0^{\rm sgh}$ where $L_0^X,
L_0^\psi, L_0^{\rm gh}$ and $L_0^{\rm sgh}$ represent contributions from
matter fields $X^\mu$, matter fields $\psi^\mu$, ghosts $b$ and $c$,
and superghosts $\beta$ and $\gamma$, respectively.  The matter part $L^{X_{0}} + L^{\psi}_{0}$ (also $\tilde L^{X}_{0} + \tilde L^{\psi}_{0}$), except for their corresponding zero-mode part which will be given latter,  is given already in (\ref{Virasoro-g-zero}).  The 
explicit expressions for the ghost part have nothing to do with T duality and can be found in any standard discussion of
superstring theories, for example in \cite{DiVecchia:1999mal},
therefore will not be presented here. The boundary states  $|B_{p} \rangle$ and  $| B_{p'} \rangle$ used above are the respective GSO projected ones given in (\ref{gsop-Dp-bs}).

In general, we have two contributions to the total amplitude given in (\ref{ampli}), one from the NS-NS sector and the other from the R-R sector. In other words, we have
\be
\Gamma_{\rm Dp'|Dp} = \Gamma^{\rm NS-NS}_{\rm Dp'|Dp} + \Gamma^{\rm R-R}_{\rm Dp'|Dp}.
\ee
Computing each of them is boiled down to the following one in each sector
\be
\Gamma_{\rm Dp'|Dp} (\eta', \eta) = \langle B^{p'}, \eta' | D |B^{p}, \eta \rangle,
\ee
with the respective boundary state given by (\ref{Dp-bs}) for which we also include the compactified case into consideration. As discussed in detail in \cite{Lu:2018suj, Jia:2019hbr},  we have $\Gamma_{\rm Dp'|Dp} (\eta', \eta) = \Gamma_{\rm Dp'|Dp} (\eta' \eta)$ and this amplitude can be factorized as
\be\label{amplitude}
\Gamma_{\rm Dp'|Dp} (\eta' \eta) =  \frac{\alpha'}{4 \pi} \int_{|z| \le 1} \frac{d^2 z}{|z|^2}
A^X \, A^{\rm bc}\, A^\psi (\eta'\eta)\, A^{\beta\gamma} (\eta'
\eta).
\ee 
In the above, we have 
\bea\label{me}
&&A^X = \langle B^{p'}_X | z^ {L^X_0} \bar z^{\tilde L^{X}_{0}}  |B^{p}_X \rangle,\quad
A^\psi (\eta' \eta) = \langle B^{p'}_\psi, \eta'| |z|^{2 L_0^\psi}
|B^{p}_\psi, \eta \rangle, \nn
&&A^{\rm bc} = \langle B_{\rm gh} | |z|^{2
L_0^{\rm gh}} | B_{\rm gh}\rangle,\quad A^{\beta\gamma} (\eta' \eta) =
\langle B_{\rm sgh}, \eta'| |z|^{2 L_0^{\rm sgh}} |B_{\rm sgh}, \eta\rangle.
\eea
As stressed earlier,  the total amplitude has a contribution from the R-R sector only when $p = p'$ for which this amplitude vanishes due to the cancellation between the contribution from the NS-NS sector and that from the R-R sector because of the 1/2 BPS nature of this system.  This certainly still holds true when a T duality is performed. We can understand this easily as follows. The bosonic zero-mode contribution to the boundary state remains the same to both sectors. The oscillator contributions to the amplitude from all sectors remain the same as before and after the T duality due to that the only quantity relevant to the T duality is the matrix $SS'^{T}$ \cite{Jia:2019hbr}, with the respective $S$ and $S'$ given by (\ref{S-matrix}) for the Dp brane and Dp$'$ brane, which remains invariant under the T duality.  As demonstrated in the previous section, the fermionic zero mode boundary state in R-R sector will have a sign change in front of $\Gamma_{11}$, for example, see the second line in (\ref{T-dual-rrzm}) or (\ref{T-dual-rrzmt}),  under the T duality. We can absorb this `-' sign by defining $\tilde\eta = - \eta$ and $\tilde\eta' = - \eta'$.  Since this zero-mode contribution to the amplitude depends only on the product $\tilde\eta \tilde\eta' = \eta\eta'$, not individual $\tilde\eta$ and $\tilde\eta'$,  therefore this contribution will also remain invariant under the T duality. 

When $p \neq p'$ for which the total amplitude is non-vanishing,  the only contribution to the amplitude now comes from the NS-NS sector. As discussed before,  only the bosonic zero-mode contribution to the amplitude will change under the T duality.  So this is our focus in computing the non-vanishing closed string tree cylinder amplitude for $p - p' = 2, 6$. The non-compactified cylinder amplitude for either $p - p' = 2$ or $p - p' = 6$ can be obtained from the general one given in \cite{Jia:2019hbr} by setting the worldvolume flux vanishing\footnote{The non-compactified amplitude for the $p - p' = 2 n$ case has also been computed in \cite{DiVecchia:1999mal} but the identities of various $\theta$ functions have not been used to simplify the integrand of the amplitude as we did in \cite{Jia:2019hbr}.}.  For $p - p' = 2$, it is for $2\le p \le 8$ 
 \be\label{t-amplitude-cylinder-p2}
\Gamma_{\rm D(p - 2)|Dp} (y)  =  \frac{ 2 \, V_{p - 1}}{(8 \pi^2 \alpha')^{\frac{p - 1}{2}}}   \int_0^\infty \frac{d t}{t^{\frac{9 - p}{2}}}  e^{- \frac{y^2}{2\pi\alpha' t}}\,\prod_{n=1}^{\infty} \frac{\left(1 + |z|^{4n}\right)^{4}}{\left(1 - |z|^{2n}\right)^{6} \left(1 + |z|^{2n}\right)^{2}}, 
\ee
where $|z| = e^{- \pi t}$ and $y$ is the brane separation along the transverse directions common to both of the branes. For $p - p' = 6$, we have the amplitude for $6 \le p \le 8$ as
\be\label{t-amplitude-cylinder-p6}
\Gamma_{\rm D(p - 6)|Dp} (y) = - \frac{ V_{p - 5} } { 2 (8 \pi^2 \alpha')^{\frac{p - 5}{2}}}  \int_0^\infty \frac{d t}{t^{\frac{9 - p}{2}}}  e^{- \frac{y^2}{2\pi\alpha' t}} 
   \prod_{n=1}^{\infty} \frac{(1 + |z|^{4n})^{4}}{(1 - |z|^{2n})^{2} (1 + |z|^{2n})^{6}},
 \ee
where again $|z| = e^{- \pi t}$ and $y$ the brane separation in the transverse directions common to both of the branes.  These two amplitudes will provide consistent checks for the corresponding compactified ones which will be computed in the following when the respective decompactification  limits are taken. Note that the infinite product in the integrand of the amplitude either (\ref{t-amplitude-cylinder-p2}) or (\ref{t-amplitude-cylinder-p6}) comes from the contribution of various oscillator modes and this will remain the same in the corresponding compactified case. So in the compactified case, we need only to compute the bosonic zero-mode contribution to the amplitude. This boils down to computing the bosonic zero-mode contribution to the matrix element of $A^{X}$ given in (\ref{me}). In other words, we need to compute
\be
A^{X}_{0} =  \,_{0}\langle B^{p'}_{X}| z^{\frac{1}{2} \alpha^{2}_{0R} + \frac{1}{2} \hat\alpha^{2}_{0}}\, {\bar z}^{\frac{1}{2} \alpha^{2}_{0L} + \frac{1}{2} \hat\alpha^{2}_{0}} |B^{p}_{X}\rangle_{0},
\ee
where $\alpha_{0R}, \alpha_{0L}$ denote the right-mover and left-mover bosonic zero modes, respectively, along the compactified directions while $\hat\alpha_{0} = \hat {\tilde\alpha}_{0}$ denote the respective bosonic zero modes along the non-compactified directions.  We will use this $A^{X}_{0}$ to replace the corresponding one in the non-compactified case in the amplitude (\ref{amplitude}) while the rest part due to contributions from various oscillators and the possible fermionic zero-mode in the R-R sector remains the same as in the non-compactified case.  Note that  in the compactified case $|B^{p}_{X} \rangle_{0} = |\Omega_{p} \rangle$, $|B^{p'}_{X} \rangle_{0} = |\Omega_{p'} \rangle$, and also the following
\be
\alpha_{0R} = \frac{l_{s}}{2} p_{R} = \frac{l_{s}}{2} \left(p_{n} - p_{\omega} \right), \quad \alpha_{0L} = \frac{l_{s}}{2} p_{L} = \frac{l_{s}}{2} \left(p_{n} + p_{\omega} \right),
\ee
and
\be
 \hat \alpha_{0} = \hat{\tilde \alpha}_{0} = \frac{l_{s}}{2} \hat p,
\ee
with $l_{s} = \sqrt{2\alpha'}$.  
So we have 
 \bea
A^{X}_{0} &=& \langle \Omega_{p'}| z^{\frac{1}{2} \alpha^{2}_{0R} + \frac{1}{2} \hat\alpha^{2}_{0}}\, {\bar z}^{\frac{1}{2} \alpha^{2}_{0L} + \frac{1}{2} \hat\alpha^{2}_{0}} |\Omega_{p} \rangle\nn
&=&  \langle \Omega_{p'}| z^{\frac{\alpha'}{4}\left[\sum_{i = 1}^{k} \left(p^{\alpha_{i}}_{n} - p_{\omega}^{\alpha_{i}}\right)^{2} +  \sum_{m = 1}^{l} \left(p^{j_{m}}_{n} - p^{j_{m}}_{\omega}\right)^{2}  + \hat p^{2}\right]} \, {\bar z}^{\frac{\alpha'}{4}\left[\sum_{i = 1}^{k} \left(p^{\alpha_{i}}_{n} + p^{\alpha_{i}}_{\omega}\right)^{2} +  \sum_{m = 1}^{l} \left(p^{j_{m}}_{n} + p^{j_{m}}_{\omega}\right)^{2}  + \hat p^{2}\right]} |\Omega_{p} \rangle\nn
&=&  \langle \Omega_{p'}| z^{\frac{\alpha'}{4}\left[\sum_{i = 1}^{k} \left( p_{\omega}^{\alpha_{i}}\right)^{2} +  \sum_{m = 1}^{l} \left(p^{j_{m}}_{n}\right)^{2}  + \hat p^{2}\right]} \, {\bar z}^{\frac{\alpha'}{4}\left[\sum_{i = 1}^{k} \left(p^{\alpha_{i}}_{\omega}\right)^{2} +  \sum_{m = 1}^{l} \left(p^{j_{m}}_{n}\right)^{2}  + \hat p^{2}\right]} |\Omega_{p} \rangle\nn
&=&  \langle \Omega_{p'}| |z|^{\frac{\alpha'}{2}\left[\sum_{i = 1}^{k} \left( p_{\omega}^{\alpha_{i}}\right)^{2} +  \sum_{m = 1}^{l} \left(p^{j_{m}}_{n}\right)^{2}  + \hat p^{2}\right]} |\Omega_{p} \rangle\nn
&=& {\cal N}_{p'} {\cal N}_{p} \langle k^{0} = 0, \hat k'^{\parallel} = 0, \hat k'^{\perp} = 0| \hat\delta^{(9 + 2 n - p -l)} (\hat q'^{\perp}) \nn
&\,& \times  \prod_{i = 1}^{k} \left[\left(\sum_{\omega'_{\alpha_{i}}} \langle  \omega'_{\alpha_{i}} |\right) |z|^{\frac{\alpha'}{2} \left( p_{\omega}^{\alpha_{i}}\right)^{2}} 
 \left(\sum_{\omega_{\alpha_{i}}} e^{ - i y^{\alpha_{i}}  \omega_{\alpha_{i}} r_{\alpha_{i}}/\alpha'} | \omega_{\alpha_{i}}\rangle\right) \right] \nn
 &\,& \times \prod_{m = 1}^{l} \left[\left(\sum_{n'_{j_{m}}} \langle n'_{j_{m}} |\right) |z|^{\frac{\alpha'}{2} \left(p^{j_{m}}_{n}\right)^{2}} \left(\sum_{n_{j_{m}}} e^{ - i y^{j_{m}} n_{j_{m}} /R_{j_{m}}} |n_{j_{m}}\rangle \right) \right] \nn
&\,& \times |z|^{\frac{\alpha'}{2} \hat p^{2} } \hat \delta^{(9 - p - l)} (\hat q^{\perp} - \hat y^{\perp}) |k^{0} = 0, \hat k^{\parallel} = 0, \hat k^{\perp} = 0 \rangle\nn
&=& {\cal N}_{p'} {\cal N}_{p}   \prod_{i = 1}^{k} \Phi_{\alpha_{i}} \left(\sum_{\omega_{\alpha_{i}}} |z|^{\frac{\omega^{2}_{\alpha_{i}} r^{2}_{\alpha_{i}}} {2 \alpha'}} e^{ - i y^{\alpha_{i}}  \omega_{\alpha_{i}} r_{\alpha_{i}}/\alpha'}\right) \prod_{m = 1}^{l} \Phi_{j_{m}} \left(\sum_{n_{j_{m}}} |z|^{\frac{\alpha' n^{2}_{j_{m}}}{2 R^{2}_{j_{m}}}} e^{ - i y^{j_{m}} n_{j_{m}} /R_{j_{m}}}\right) \nn
&&\times   \langle k^{0} = 0, \hat k'^{\parallel} = 0, \hat k'^{\perp} = 0| \hat\delta^{(9 + 2 n - p - l)} (\hat q'^{\perp}) |z|^{\frac{\alpha'}{2} \hat p^{2} } \hat \delta^{(9 - p - l )} (\hat q^{\perp} - \hat y^{\perp}) \nn
&&\times |k^{0} = 0, \hat k^{\parallel} = 0, \hat k^{\perp} = 0 \rangle,
\eea
where in the third line we have used $p^{\alpha_{i}}_{n} |\Omega_{p} \rangle = 0, p^{j_{m}}_{\omega} |\Omega_{p} \rangle = 0$ as given in (\ref{momentum-eigen}) and in the fifth line we have used the explicit expression for $|\Omega_{p} \rangle$ given in (\ref{compactified-pbs}). Note also that $|y^{j_{m}}|$ is the brane separation along the respective compactified direction. In the following, we move to compute the factor in the last two lines above. For this, note that 
\be
V_{1 + p - 2n - k} =  \langle k^{0} = 0, \hat k'^{\parallel} = 0|k^{0} = 0, \hat k'^{\parallel} = 0 \rangle,
\ee
we have then

\bea
&& \langle k^{0} = 0, \hat k'^{\parallel} = 0, \hat k'^{\perp} = 0| \hat\delta^{(9 + 2 n - p - l)} (\hat q'^{\perp}) |z|^{\frac{\alpha'}{2} \hat p^{2} } \hat \delta^{(9 - p - l)} (\hat q^{\perp} - \hat y^{\perp}) |k^{0} = 0, \hat k^{\parallel} = 0, \hat k^{\perp} = 0 \rangle\nn
&=& \int \frac{d^{2n + 9 - p - l} Q'}{(2 \pi)^{2n + 9 - p - l}} \int \frac{d^{9 - p - l} Q}{(2 \pi)^{9 - p - l}} \langle k^{0} = 0, \hat k'^{\parallel} = 0, \hat k'^{\perp} = - Q'|  |z|^{\frac{\alpha'}{2} \hat p^{2} } e^{- i Q\cdot \hat y^{\perp}}\nn
&&\quad \times  |k^{0} = 0, \hat k^{\parallel} = 0, \hat k^{\perp} = Q \rangle\nn
&= &  V_{1 + 2 p - 2n - k} \int d^{2n + 9 - p - l} Q' \int \frac{d^{9 - p - l} Q}{(2 \pi)^{9 - p - l}}  |z|^{\frac{\alpha'}{2} Q^{2} } e^{- i Q\cdot \hat y^{\perp}} \delta^{(2n)} (Q')  \delta^{(9 - p - l)} (Q + Q')\nn
&=& V_{1 + 2 p - 2n - k}  \int \frac{d^{9 - p - l} Q}{(2 \pi)^{9 - p - l}}  |z|^{\frac{\alpha'}{2} Q^{2} } e^{- i Q\cdot \hat y^{\perp}}.  
 \eea 
 So we have
 \bea
 &&A^{X}_{0} = {\cal N}_{p'} {\cal N}_{p}   \prod_{i = 1}^{k} \Phi_{\alpha_{i}} \left(\sum_{\omega_{\alpha_{i}}} |z|^{\frac{\omega^{2}_{\alpha_{i}} r^{2}_{\alpha_{i}}} {2 \alpha'}} e^{ - i y^{\alpha_{i}}  \omega_{\alpha_{i}} r_{\alpha_{i}}/\alpha'}\right) \prod_{m = 1}^{l} \Phi_{j_{m}} \left(\sum_{n_{j_{m}}} |z|^{\frac{\alpha' n^{2}_{j_{m}}}{2 R^{2}_{j_{m}}}} e^{ - i y^{j_{m}} n_{j_{m}} /R_{j_{m}}}\right)\nn
 &&\quad  \times V_{1 + 2 p - 2n - k}  \int \frac{d^{9 - p - l} Q}{(2 \pi)^{9 - p - l}}  |z|^{\frac{\alpha'}{2} Q^{2} } e^{- i Q\cdot \hat y^{\perp}}\nn 
  &&=   {\cal N}_{p'} {\cal N}_{p}  \prod_{i = 1}^{k} \prod_{m = 1}^{l}  \Phi_{\alpha_{i}} \Phi_{j_{m}} \left(\sum_{\omega_{\alpha_{i}}} e^{- \frac{  \pi t\, \omega^{2}_{\alpha_{i}} r^{2}_{\alpha_{i}}} {2 \alpha'} - i y^{\alpha_{i}}  \omega_{\alpha_{i}} r_{\alpha_{i}}/\alpha'}\right) 
  \left(\sum_{n_{j_{m}}} e^{-\frac{ \pi t \alpha' n^{2}_{j_{m}}}{2 R^{2}_{j_{m}}}  - i y^{j_{m}} n_{j_{m}} /R_{j_{m}}} \right) \nn
  &&\quad \times V_{1 + 2 p - 2n - k}  \int \frac{d^{9 - p - l} Q}{(2 \pi)^{9 - p - l}}  e^{- \frac{ \pi t \alpha'}{2} Q^{2} - i Q\cdot \hat y^{\perp}},
 \eea 
 where we have set $|z| = e^{- \pi t}$.  Note that
 \bea
 {\cal N}_{p'} {\cal N}_{p} \prod_{i = 1}^{k} \prod_{m = 1}^{l}  \Phi_{\alpha_{i}} \Phi_{j_{m}} &=& \frac{c_{p - 2n} c_{p}}{4}  \prod_{i = 1}^{k} \frac{2 \pi r_{\alpha_{i}}}{\Phi_{\alpha_{i}}} \prod_{m = 1}^{l} \frac{1}{2 \pi R_{j_{m}} \Phi_{j_{m}}}  \prod_{i = 1}^{k} \prod_{m = 1}^{l}  \Phi_{\alpha_{i}} \Phi_{j_{m}} \nn
 &=&  \frac{c_{p - 2n} c_{p}}{4} \prod_{i = 1}^{k} 2\pi r_{\alpha_{i}} \prod_{m = 1}^{l} \frac{1}{2 \pi R_{j_{m}}},
  \eea 
where we have used (\ref{normalization}) for ${\cal N}_{p}$ and ${\cal N}_{p'}$, and
\be
\int \frac{d^{9 - p - l} Q}{(2 \pi)^{9 - p - l}}  e^{- \frac{ \pi t \alpha'}{2} Q^{2} - i Q\cdot \hat y^{\perp}} = (2 \pi^{2} \alpha' t)^{- \frac{9 - p -l}{2}} e^{- \frac{(\hat y^{\perp})^{2}}{2\pi \alpha' t}}.
\ee 
 With the above, we have
 \bea
 A^{X}_{0} &=&  \frac{c_{p - 2n} c_{p}}{4 (2 \pi^{2} \alpha')^{\frac{9 - p - l}{2}}} V_{1 + p - 2n - k} \prod_{i = 1}^{k} \left(2 \pi r_{\alpha_{i}} \sum_{\omega_{\alpha_{i}}} e^{- \frac{  \pi t\, \omega^{2}_{\alpha_{i}} r^{2}_{\alpha_{i}}} {2 \alpha'} - i \frac{y^{\alpha_{i}}  \omega_{\alpha_{i}} r_{\alpha_{i}}}{\alpha'}}\right) \nn
 &\,& \times  \prod_{m = 1}^{l} \left(\frac{1}{2 \pi R_{j_{m}}} \sum_{n_{j_{m}}} e^{-\frac{ \pi t \alpha' n^{2}_{j_{m}}}{2 R^{2}_{j_{m}}}  - i \frac{y^{j_{m}} n_{j_{m}}}{R_{j_{m}}}} \right)  \frac{e^{- \frac{(\hat y^{\perp})^{2}}{2\pi\alpha' t}}} {t^{\frac{9 - p -l}{2}}}.
 \eea 
 Let us have a consistent check of the above to see if it gives the decompactification limit when we take $r_{\alpha_{i}} \to \infty$ and $R_{j_{m}} \to \infty$.  When we take $r_{\alpha_{i}} \to \infty$, it is clear that only the winding $\omega_{\alpha_{i}} = 0$ term survives. When $R_{j_{m}} \to \infty$ is taken, we set $k_{j_{m}} = n_{j_{m}} /R_{j_{m}} \to d k_{j_{m}} = 1/R_{j_{m}}$ and pass the summation to an integration as follows
\be
 \sum_{n_{j_{m}}}  e^{-\frac{ \pi t \alpha' n^{2}_{j_{m}}}{2 R^{2}_{j_{m}}}  - i \frac{y^{j_{m}} n_{j_{m}}}{R_{j_{m}}}} = R_{j_{m}} \int d k_{j_{m}} e^{-\frac{ \pi t \alpha' k^{2}_{j_{m}}}{2}  - i y^{j_{m}} k_{j_{m}}} = 2 \pi R_{j_{m}} \left(2 \pi^{2}\alpha' t\right)^{- 1/2} e^{- \frac{(y^{j_{m}})^{2}}{2\pi \alpha' t}}.
 \ee
 We then have
 \bea
 A^{X}_{0} &\to &  \frac{c_{p - 2n} c_{p}}{4 (2 \pi^{2} \alpha')^{\frac{9 - p - l}{2}}} V_{1 + p - 2n - k} \left(\prod_{i = 1}^{k} 2 \pi r_{\alpha_{i}}  \right) \left(\prod_{m = 1}^{l} \frac{e^{- \frac{(y^{j_{m}})^{2}}{2\pi \alpha' t}}}{\sqrt{2\pi^{2} \alpha' t}} \right) \frac{e^{- \frac{(\hat y^{\perp})^{2}}{2\pi\alpha' t}}} {t^{\frac{9 - p -l}{2}}}\nn
 &=& \frac{c_{p - 2n} c_{p} V_{1 + p - 2n}}{4 (2 \pi^{2} \alpha')^{\frac{9 - p }{2}}} \frac{e^{- \frac{(\hat y^{\perp})^{2} + \sum_{m = 1}^{l} \left( y^{j_{m}}\right)^{2}} {2\pi\alpha' t}}} {t^{\frac{9 - p }{2}}}, 
 \eea
where we have set $V_{1 + p - 2n} = V_{1 + p + 2n -k} \prod_{i = 1}^{k} 2 \pi r_{\alpha_{i}}$ with $r_{\alpha_{i}} \to \infty$.   Note also that $y^{2} = (\hat y^{\perp})^{2} + \sum_{m = 1}^{l} (y^{j_{m}})^{2}$ is precisely the brane separation along the transverse directions to both branes when the decompactification limit is taken.  Therefore
\be
A^{X}_{0} \to  \frac{c_{p - 2n} c_{p} V_{1 + p - 2n}}{4 (2 \pi^{2} \alpha')^{\frac{9 - p }{2}}} \frac{e^{- \frac{y^{2}} {2\pi\alpha' t}}} {t^{\frac{9 - p }{2}}},
\ee
giving precisely the corresponding decompactification limit. 

With the above, we have the compactified cylinder amplitude between a D(p - 2n) brane and a Dp brane as
\bea\label{compactified-cylinder-ampli}
\Gamma_{\rm D(p - 2n)|Dp}  &=&  \frac{c_{p - 2n} c_{p}}{4}  \frac{\alpha'}{4\pi} 2 \pi^{2} \frac{2^2\,V_{1 + p - 2 n - k}}{(2 \pi^{2} \alpha')^{\frac{9 - p - l}{2}}} \int_{0}^{\infty} \frac{d t} {t^{\frac{9 - p -l}{2}}} e^{- \frac{(\hat y^{\perp})^{2}}{2\pi\alpha' t}} \,Z_{\rm (p - 2n) | p} (t)\nn
&\times& \prod_{i = 1}^{k} \left[2 \pi r_{\alpha_{i}} \sum_{\omega_{\alpha_{i}}} e^{- \frac{  \pi t\, \omega^{2}_{\alpha_{i}} r^{2}_{\alpha_{i}}} {2 \alpha'} - i \frac{y^{\alpha_{i}}  \omega_{\alpha_{i}} r_{\alpha_{i}}}{\alpha'}}\right]   \prod_{m = 1}^{l} \left[\frac{1}{2 \pi R_{j_{m}}} \sum_{n_{j_{m}}} e^{-\frac{ \pi t \alpha' n^{2}_{j_{m}}}{2 R^{2}_{j_{m}}}  - i \frac{y^{j_{m}} n_{j_{m}}}{R_{j_{m}}}} \right],\nn 
\eea 
 where we have used 
 \be
 \int_{|z|\le 1} \frac{d z^{2}}{|z|^{2}} = 2 \pi^{2} \int_{0}^{\infty} d t.
 \ee
As pointed out early,  the various oscillator contributions plus possible fermionic zero-mode contribution in the R-R sector to the amplitude remain inert with respect to the compactifications and we denote these contributions by  $ 2^{2} Z_{\rm (p - 2n)|p}$ in the above amplitude.  Note that $Z_{\rm (p - 2n)|p} (t) = 0$ for $n = 0, 2$ due to the BPS nature of the underlying systems, for $n = 1$ from (\ref{t-amplitude-cylinder-p2})
\be
Z_{\rm (p - 2)|p} (t) = \frac{\left(1 + |z|^{4n}\right)^{4}}{\left(1 - |z|^{2n}\right)^{6} \left(1 + |z|^{2n}\right)^{2}},
\ee
and for $n = 3$ from (\ref{t-amplitude-cylinder-p6})
\be
Z_{\rm (p - 6)|p} (t) =  \prod_{n=1}^{\infty} \frac{(1 + |z|^{4n})^{4}}{(1 - |z|^{2n})^{2} (1 + |z|^{2n})^{6}}.
\ee
Note that
\be
c_{p} = \sqrt{\pi} \left(2 \pi \sqrt{\alpha'}\right)^{3 - p}, \quad c_{p - 2n} = \sqrt{\pi} \left(2\pi \sqrt{\alpha'}\right)^{3 + 2n - p},
\ee
we have
\be
c_{p} c_{p - 2n} \frac{\alpha'}{4 \pi} 2 \pi^{2} = \frac{2^{n}}{2^{p - 1}} \left(2 \pi^{2} \alpha'\right)^{4 + n - p}.
\ee
So we have the compactifid cylinder amplitude (\ref{compactified-cylinder-ampli}) as
\bea\label{compactified-cylinder-amplitude}
\Gamma_{\rm D(p - 2n)|Dp} (y) &=& \frac{ 2^{2 - n - l} \, V_{1 + p - 2n - k}}{(8 \pi^{2} \alpha')^{\frac{1 + p - 2 n - l}{2}}} \int_{0}^{\infty} \frac{d t} {t^{\frac{9 - p -l}{2}}} e^{- \frac{(\hat y^{\perp})^{2}}{2\pi\alpha' t}} \,Z_{\rm (p - 2n) | p} (t)\nn
&\times& \prod_{i = 1}^{k} \left[2 \pi r_{\alpha_{i}} \sum_{\omega_{\alpha_{i}}} e^{- \frac{  \pi t\, \omega^{2}_{\alpha_{i}} r^{2}_{\alpha_{i}}} {2 \alpha'} - i \frac{y^{\alpha_{i}}  \omega_{\alpha_{i}} r_{\alpha_{i}}}{\alpha'}}\right]   \prod_{m = 1}^{l} \left[\frac{1}{2 \pi R_{j_{m}}} \sum_{n_{j_{m}}} e^{-\frac{ \pi t \alpha' n^{2}_{j_{m}}}{2 R^{2}_{j_{m}}}  - i \frac{y^{j_{m}} n_{j_{m}}}{R_{j_{m}}}} \right].\nn
\eea 
Taking the decompactification limit, as discussed earlier, we have
\bea\label{decomp-limit}
&&\prod_{i = 1}^{k} \left[2 \pi r_{\alpha_{i}} \sum_{\omega_{\alpha_{i}}} e^{- \frac{  \pi t\, \omega^{2}_{\alpha_{i}} r^{2}_{\alpha_{i}}} {2 \alpha'} - i \frac{y^{\alpha_{i}}  \omega_{\alpha_{i}} r_{\alpha_{i}}}{\alpha'}}\right] \to \prod_{i = 1}^{k} 2 \pi r_{\alpha_{i}}, \nn
&&  \prod_{m = 1}^{l} \left[\frac{1}{2 \pi R_{j_{m}}} \sum_{n_{j_{m}}} e^{-\frac{ \pi t \alpha' n^{2}_{j_{m}}}{2 R^{2}_{j_{m}}}  - i \frac{y^{j_{m}} n_{j_{m}}}{R_{j_{m}}}} \right] \to \prod_{m = 1}^{l} \frac{e^{- \frac{(y^{j_{m}})^{2}}{2\pi \alpha' t}}}{\sqrt{2\pi^{2} \alpha' t}} = \frac{e^{- \sum_{m = 1}^{l} \frac{(y^{j_{m}})^{2}}{2\pi \alpha' t}}}{(2\pi^{2} \alpha' t)^{\frac{l}{2}}}.
\eea
We then have the decompactified cylinder amplitude from (\ref{compactified-cylinder-amplitude}) as
\be \label{t-amplitude-cylinder-p2n}
\Gamma_{\rm D(p - 2 n)|Dp} (y) =  \frac{ 2^{2 - n} \, V_{1 + p - 2n}}{(8 \pi^{2} \alpha')^{\frac{1 + p - 2 n}{2}}} \int_{0}^{\infty} \frac{d t} {t^{\frac{9 - p}{2}}} e^{- \frac{ y^{2}}{2\pi\alpha' t}} \,Z_{\rm (p - 2n) | p} (t),
\ee
where as before $V_{1 + p - 2 n} = V_{1 + p - 2n - k} \prod_{i = 1}^{k} 2 \pi r_{\alpha_{i}}$ for $r_{\alpha_{i}} \to \infty$ and $y^{2} = (\hat y^{\perp})^{2} + \sum_{m = 1}^{l} (y^{j_{m}})^{2}$.
This gives precisely the one in (\ref{t-amplitude-cylinder-p2}) for $n = 1$ and the one in (\ref{t-amplitude-cylinder-p6}) for $n = 3$.

In summary, we have in this section computed the compactified cylinder amplitude (\ref{compactified-cylinder-amplitude}) between a D(p - 2n) and a Dp with $2n \le p \le 8$. This amplitude has $k \le p - 2 n$ compactified longitudinal directions and $l \le 9 - p$ compactified transverse directions common to both the branes.  As specified earlier, the respective compactfified radii are $r_{\alpha_{i}}$ and $R_{j_{m}}$ with $i = 1, 2, \cdots k$ and $m = 1, 2, \cdots l$.  These two D branes are placed parallel along the non-compactified transverse directions at a separation $\hat y^{\perp}$ and along each of the compactified transverse directions at $|y^{j_{m}}|$.  $y^{\alpha_{i}}$ are the Wilson lines turned on along the respective compactified worldvolume directions of the Dp brane while keeping the D(p - 2n) brane absent of these. 

\section{T duality}
Given the cylinder amplitude (\ref{compactified-cylinder-amplitude}), performing a T duality along either a compactified longitudinal direction or a compactified transverse direction becomes easy. 

Without loss of generality, let us first perform this T duality specifically along the longitudinal $\alpha_{k}$ direction. We expect to obtain the corresponding compactified amplitude from $\Gamma_{\rm D(p - 2n)|Dp} \to \Gamma_{\rm D(p - 2n - 1|D(p - 1)}$.  Let us check if this is indeed true.  For this, we send $r_{\alpha_{k}} \to R_{j_{l + 1}} = \alpha'/r_{\alpha_{k}}$, $\omega_{\alpha_{k}} \to n_{j_{l + 1}} = \omega_{\alpha_{k}}$,  and $y^{\alpha_{k}} \to y^{j_{l + 1}} = y^{\alpha_{k}}$  to the cylinder amplitude (\ref{compactified-cylinder-amplitude}).  We have then 
\be
2 \pi r_{\alpha_{k}} \sum_{\omega_{\alpha_{k}}} e^{- \frac{  \pi t\, \omega^{2}_{\alpha_{k}} r^{2}_{\alpha_{k}}} {2 \alpha'} - i \frac{y^{\alpha_{k}}  \omega_{\alpha_{k}} r_{\alpha_{k}}}{\alpha'}}
\to \frac{8 \pi^{2} \alpha'}{2} \frac{1}{2 \pi R_{j_{l + 1}}} \sum_{n_{j_{l + 1}}}  e^{- \frac{  \pi t \alpha' \, n^{2}_{j_{l + 1}} } {2 R^{2}_{j_{l + 1}}} - i \frac{y^{j_{l + 1}} n_{j_{l + 1}}}{R_{j_{l + 1}}}}.
\ee
We have then from  (\ref{compactified-cylinder-amplitude})
\bea
\Gamma_{\rm D(p - 2n)|Dp} &\to&  \frac{ 2^{2 - n - l} \, V_{1 + p - 2n - k}}{(8 \pi^{2} \alpha')^{\frac{1 + p - 2 n - l}{2}}} \int_{0}^{\infty} \frac{d t} {t^{\frac{9 - p -l}{2}}} e^{- \frac{(\hat y^{\perp})^{2}}{2\pi\alpha' t}} \,Z_{\rm (p - 2n - 1) | (p - 1)} (t)\nn
&\times& \prod_{i = 1}^{k- 1} \left[2 \pi r_{\alpha_{i}} \sum_{\omega_{\alpha_{i}}} e^{- \frac{  \pi t\, \omega^{2}_{\alpha_{i}} r^{2}_{\alpha_{i}}} {2 \alpha'} - i \frac{y^{\alpha_{i}}  \omega_{\alpha_{i}} r_{\alpha_{i}}}{\alpha'}}\right]   \prod_{m = 1}^{l} \left[\frac{1}{2 \pi R_{j_{m}}} \sum_{n_{j_{m}}} e^{-\frac{ \pi t \alpha' n^{2}_{j_{m}}}{2 R^{2}_{j_{m}}}  - i \frac{y^{j_{m}} n_{j_{m}}}{R_{j_{m}}}} \right]\nn
&\times&  \frac{8 \pi^{2} \alpha'}{2} \frac{1}{2 \pi R_{j_{l + 1}}} \sum_{n_{j_{l + 1}}}  e^{- \frac{  \pi t \alpha' \, n^{2}_{j_{l + 1}} } {2 R^{2}_{j_{l + 1}}} - i \frac{y^{j_{l + 1}} n_{j_{l + 1}}}{R_{j_{l + 1}}}}\nn
&=&  \frac{ 2^{2 - n - (l + 1)} \, V_{1 + (p - 1) - 2n - (k - 1)}}{(8 \pi^{2} \alpha')^{\frac{1 + (p - 1) - 2 n - (l + 1)}{2}}} \int_{0}^{\infty} \frac{d t} {t^{\frac{9 - (p - 1) - (l + 1)}{2}}} e^{- \frac{(\hat y^{\perp})^{2}}{2\pi\alpha' t}} \,Z_{\rm (p - 2n - 1) | (p - 1)} (t)\nn
&\times& \prod_{i = 1}^{k- 1} \left[2 \pi r_{\alpha_{i}} \sum_{\omega_{\alpha_{i}}} e^{- \frac{  \pi t\, \omega^{2}_{\alpha_{i}} r^{2}_{\alpha_{i}}} {2 \alpha'} - i \frac{y^{\alpha_{i}}  \omega_{\alpha_{i}} r_{\alpha_{i}}}{\alpha'}}\right]   \prod_{m = 1}^{l + 1} \left[\frac{1}{2 \pi R_{j_{m}}} \sum_{n_{j_{m}}} e^{-\frac{ \pi t \alpha' n^{2}_{j_{m}}}{2 R^{2}_{j_{m}}}  - i \frac{y^{j_{m}} n_{j_{m}}}{R_{j_{m}}}} \right]\nn
&=& \Gamma_{\rm D(p - 2n -1)|D(p - 1)},
\eea
where in the first line we have sent  $Z_{\rm (p - 2n)|p} (t) \to Z_{\rm (p - 2n - 1)|(p - 1)} (t) =  Z_{\rm (p - 2n)|p} (t)$ due to the inert of various oscillator contributions to the amplitude under T duality and $\hat y^{\perp}$ remains the same.  So this goes indeed as expected. 

We now move to discuss performing a T duality along a compactified transverse direction, say, along $j_{l}$, to the amplitude (\ref{compactified-cylinder-amplitude}). We then expect to have $\Gamma_{\rm D(p - 2n)|Dp} \to \Gamma_{\rm D(p + 1 - 2n)|D (p + 1)}$.  For this, we send 
$R_{j_{l}} \to r_{\alpha_{k + 1}} = \alpha'/R_{j_{l}}$, $n_{j_{l}} \to \omega_{\alpha_{k +1}} = n_{j_{l}}$ and $y^{j_{l}} \to y^{\alpha_{k + 1}} = y^{j_{l}}$.  We  have 
\be
\frac{1}{2 \pi R_{j_{m}}} \sum_{n_{j_{m}}} e^{-\frac{ \pi t \alpha' n^{2}_{j_{m}}}{2 R^{2}_{j_{m}}}  - i \frac{y^{j_{m}} n_{j_{m}}}{R_{j_{m}}}} \to \frac{2}{8 \pi^{2} \alpha'}  2 \pi r_{\alpha_{k + 1}}\sum_{\omega_{\alpha_{k + 1}}} e^{-\frac{ \pi t  \omega^{2}_{\alpha_{k + 1}} r^{2}_{\alpha_{k + 1}}}{2 \alpha'}  - i \frac{y^{\alpha_{k + 1}} \omega_{\alpha_{k + 1}} r_{\alpha_{k + 1}}}{\alpha'}}.
\ee
We have then from (\ref{compactified-cylinder-amplitude}) 
\bea
\Gamma_{\rm D(p - 2 n)|Dp} &\to&  \frac{ 2^{2 - n - l} \, V_{1 + p - 2n - k}}{(8 \pi^{2} \alpha')^{\frac{1 + p - 2 n - l}{2}}} \int_{0}^{\infty} \frac{d t} {t^{\frac{9 - p -l}{2}}} e^{- \frac{(\hat y^{\perp})^{2}}{2\pi\alpha' t}} \,Z_{\rm (p + 1 - 2n) | (p + 1)} (t)\nn
&\times& \prod_{i = 1}^{k} \left[2 \pi r_{\alpha_{i}} \sum_{\omega_{\alpha_{i}}} e^{- \frac{  \pi t\, \omega^{2}_{\alpha_{i}} r^{2}_{\alpha_{i}}} {2 \alpha'} - i \frac{y^{\alpha_{i}}  \omega_{\alpha_{i}} r_{\alpha_{i}}}{\alpha'}}\right]   \prod_{m = 1}^{l - 1} \left[\frac{1}{2 \pi R_{j_{m}}} \sum_{n_{j_{m}}} e^{-\frac{ \pi t \alpha' n^{2}_{j_{m}}}{2 R^{2}_{j_{m}}}  - i \frac{y^{j_{m}} n_{j_{m}}}{R_{j_{m}}}} \right]\nn
&\times& \frac{2}{8 \pi^{2} \alpha'} 2 \pi r_{\alpha_{k + 1}} \sum_{\omega_{\alpha_{k + 1}}} e^{-\frac{ \pi t  \omega^{2}_{\alpha_{k + 1}} r^{2}_{\alpha_{k + 1}}}{2 \alpha'}  - i \frac{y^{\alpha_{k + 1}} \omega_{\alpha_{k + 1}} r_{\alpha_{k + 1}}}{\alpha'}}\nn
&=&  \frac{ 2^{2 - n - (l - 1)} \, V_{1 + (p + 1) - 2n - (k + 1)}}{(8 \pi^{2} \alpha')^{\frac{1 + (p + 1) - 2 n - (l - 1)}{2}}} \int_{0}^{\infty} \frac{d t} {t^{\frac{9 - (p + 1) - (l - 1)}{2}}} e^{- \frac{(\hat y^{\perp})^{2}}{2\pi\alpha' t}} \,Z_{\rm (p + 1 - 2n) | (p + 1)} (t)\nn
&\times& \prod_{i = 1}^{k + 1} \left[2 \pi r_{\alpha_{i}} \sum_{\omega_{\alpha_{i}}} e^{- \frac{  \pi t\, \omega^{2}_{\alpha_{i}} r^{2}_{\alpha_{i}}} {2 \alpha'} - i \frac{y^{\alpha_{i}}  \omega_{\alpha_{i}} r_{\alpha_{i}}}{\alpha'}}\right]   \prod_{m = 1}^{l - 1} \left[\frac{1}{2 \pi R_{j_{m}}} \sum_{n_{j_{m}}} e^{-\frac{ \pi t \alpha' n^{2}_{j_{m}}}{2 R^{2}_{j_{m}}}  - i \frac{y^{j_{m}} n_{j_{m}}}{R_{j_{m}}}} \right]\nn
&=& \Gamma_{\rm D(p + 1 - 2 n)|D (p + 1)},
\eea
where in the first line we have also set $Z_{\rm (p + 1 - 2n)|(p + 1)} (t) = Z_{\rm (p - 2n)|Dp} (t)$ for the same reason as explained earlier.  Indeed we obtain the expected one.

As demonstrated already in the previous section, the decompactification limit will give the respective known cylinder amplitude after the T dualities discussed above. 

In the following, we will explain, as promised in Introduction, that the usual low energy approach  does work to a T duality along a transverse direction common to the two D branes for the corresponding decompactified cylinder amplitude but does not work to a T duality along a longitudinal direction. 

The simple reason is that to obtain the decompactified result after the T duality, we need to send the initial compactified radius vanishing. To perform a low energy T duality along a transverse direction, we first need to generate an isometry along this direction by periodically placed this system along this direction and by sending the radius vanishing.  In other words, we don't consider the momentum modes along this effective compactification with a zero radius. So we don't have a winding in the T dual system which is consistent with the fact in that the T dual radius becomes infinite large and there is no winding allowed. However, this is not the story if we do a low energy T dual along a longitudinal direction.  The system has already an isometry along the longitudinal direction since the decompactified amplitude is independent of the longitudinal directions. If we simply assume one of them compactified with a vanishing radius, it is clear that an infinite number of windings is not accounted, opposite to what we did in (\ref{compactified-cylinder-amplitude}).  This implies that in the T dual system with an infinite radius, the contribution from the corresponding infinite number of momentum modes which gives a continuous momentum in the infinite large radius limit to the amplitude has not been accounted for\footnote{This contribution gives rise to one of the factors given in the second equation in (\ref{decomp-limit}).}. So this kind of T duality does not work to the decomplactified amplitude if it is performed along a longitudinal direction. 

We here give an explicit demonstration of the T duality acting on the non-compactified or the decompactified cylinder amplitude (\ref{t-amplitude-cylinder-p2n}) along a transverse direction to both the branes.
For this, we first write $y^{2} = y^{2}_{p + 1} + y^{2}_{p + 2} + \cdots y^{2}_{9} = y^{2}_{p + 1} + \bar y^{2}$. We will perform a T duality along $y_{p + 1}$-direction. For this, we first need to place the ${\rm D(p - 2n)|Dp}$  system periodically along this direction to an effective compactification along this direction.  This can be done by placing D(p - 2n) and Dp as 
\be
\vec{y}_{\rm D(p - 2 n)} = 2 \pi k a \,\hat e_{p + 1}, \quad k = 0, \pm 1, \pm 2, \cdots,
\ee
and
\be
\vec{y}_{\rm Dp} = \vec{\bar y} + (y_{p + 1} + 2 \pi l a)\, \hat e_{p + 1}, \quad l = 0, \pm 1, \pm 2, \cdots,
\ee
where $\hat e_{p + 1}$ is the unit vector along $y_{p + 1}$ and $a$ is the radius of compactification with $a \to 0$. We then have
\be
\vec{y}_{k, l} = \vec{y}_{\rm Dp} - \vec{y}_{\rm D(p - 2n)} = \vec{\bar y} + \left[y_{p + 1} + 2 \pi a (l - k)\right] \hat e_{p + 1}.
\ee
So we have now
\be
y^{2}_{k, l} = \bar y^{2} + \left[y_{p + 1} +  2 \pi a (l - k)\right]^{2}.
\ee
Then the total cylinder amplitude between the periodic D(p - 2n) and the periodic Dp with $\Gamma^{(k, l)}_{\rm D(p - 2n)|Dp}$ given by (\ref{decomp-limit}) is
\bea
\Gamma &=& \sum_{k, l} \Gamma^{(k, l)}_{\rm D(p - 2 n)|Dp} (y_{k, l})\nn
&=&   \frac{ 2^{2 - n} \, V_{1 + p - 2n}}{(8 \pi^{2} \alpha')^{\frac{1 + p - 2 n}{2}}} \int_{0}^{\infty} \frac{d t} {t^{\frac{9 - p}{2}}}  \,Z_{\rm (p - 2n) | p} (t) \, \sum_{k, l} e^{- \frac{ y_{k, l}^{2}}{2\pi\alpha' t}}.
\eea 
As stressed earlier, the current approach is valid only when we take the compactified radius $a \to 0$ since the momentum modes along the compactified circle has not been taken into consideration.  Note that in the above integrand, we have the sum
\be 
\sum_{k, l} e^{- \frac{ y_{k, l}^{2}}{2\pi\alpha' t}} =  \sum_{k, l} e^{- \frac{ \bar y^{2} + \left[y_{p + 1} +  2 \pi a (l - k)\right]^{2}}{2 \pi \alpha' t}} = e^{- \frac{\bar y^{2}}{2\pi \alpha' t}} \sum_{k, l} e^{- \frac{\left[y_{p + 1} +  2 \pi a (l - k)\right]^{2}}{2 \pi \alpha' t}}.
\ee
If we set $m = l - k$, the above sum can be expressed as
\be
\sum_{k, l} e^{- \frac{ y_{k, l}^{2}}{2\pi\alpha' t}} = N_{\infty}\,  e^{- \frac{\bar y^{2}}{2\pi \alpha' t}} \sum_{m}\, e^{- \frac{\left(y_{p + 1} +  2 \pi a m \right)^{2}}{2 \pi \alpha' t}},
\ee
where $N_{\infty} = \sum_{k}$.  We now set $z = y_{p + 1} + 2 \pi a m$ and $d z = 2\pi a \to 0$ if $a \to 0$. So we can pass
\be
\sum_{m} e^{- \frac{\left(y_{p + 1} +  2 \pi a m \right)^{2}}{2 \pi \alpha' t}} = \int_{ - \infty}^{\infty} \frac{d z}{2 \pi a} e^{- \frac{z^{2}}{2 \pi \alpha' t}} = \frac{\sqrt{2 \pi^{2} \alpha' t}}{2 \pi a}.
\ee
We then have the total amplitude
\be
\Gamma = N_{\infty} \frac{(2 \pi^{2} \alpha')^{1/2}}{2 \pi a} \frac{ 2^{2 - n} \, V_{1 + p - 2n}}{(8 \pi^{2} \alpha')^{\frac{1 + p - 2 n}{2}}} \int_{0}^{\infty} \frac{d t} {t^{\frac{8 - p}{2}}} \, e^{- \frac{\bar y^{2}}{2\pi \alpha' t}}  \,Z_{\rm (p - 2n) | p} (t). 
\ee
So the cylinder amplitude between a D(p - 2n) and a Dp with such a transverse compactification with $a \to 0$ is given as
\bea
\Gamma_{\rm D(p - 2n)|Dp} (\bar y) &=& \frac{\Gamma}{N_{\infty}} = \frac{(2 \pi^{2} \alpha')^{1/2}}{2 \pi a} \frac{ 2^{2 - n} \, V_{1 + p - 2n}}{(8 \pi^{2} \alpha')^{\frac{1 + p - 2 n}{2}}} \int_{0}^{\infty} \frac{d t} {t^{\frac{8 - p}{2}}} \, e^{- \frac{\bar y^{2}}{2\pi \alpha' t}}  \,Z_{\rm (p - 2n) | p} (t)\nn
&=&  \frac{1}{2 \pi a} \frac{ 2^{2 - n - 1} \, V_{1 + p - 2n}}{(8 \pi^{2} \alpha')^{\frac{1 + p - 2 n - 1}{2}}} \int_{0}^{\infty} \frac{d t} {t^{\frac{8 - p}{2}}} \, e^{- \frac{\bar y^{2}}{2\pi \alpha' t}}  \,Z_{\rm (p - 2n) | p} (t),
\eea
which corresponds precisely to the $l = 1, k = 0$ case, when $a \to 0$, given in (\ref{compactified-cylinder-amplitude}) in the previous section.  This also clearly shows that the low energy T duality only works when $a \to 0$ (otherwise the momentum modes must be taken into consideration).  T duality along this direction amounts to sending $a \to \bar a = \alpha'/a \to \infty$ in the above for which we have
\bea
\Gamma_{\rm D(p - 2n)|Dp} &\to&  \frac{\bar a}{2 \pi \alpha'} \frac{ 2^{2 - n - 1} \, V_{1 + p - 2n}}{(8 \pi^{2} \alpha')^{\frac{1 + p - 2 n - 1}{2}}} \int_{0}^{\infty} \frac{d t} {t^{\frac{8 - p}{2}}} \, e^{- \frac{\bar y^{2}}{2\pi \alpha' t}}  \,Z_{\rm (p - 2n) | p} (t)\nn
&=& \frac{ 2 \pi \bar a}{4 \pi^{2} \alpha'} \frac{ 2^{2 - n - 1} \, V_{1 + p - 2n}}{(8 \pi^{2} \alpha')^{\frac{1 + p - 2 n - 1}{2}}} \int_{0}^{\infty} \frac{d t} {t^{\frac{8 - p}{2}}} \, e^{- \frac{\bar y^{2}}{2\pi \alpha' t}}  \,Z_{\rm (p + 1 - 2n) | (p + 1)} (t)\nn
&=& \frac{ 2^{2 - n} \, V_{1 + (p + 1)  - 2n}}{(8 \pi^{2} \alpha')^{\frac{1 + (p + 1) - 2 n }{2}}} \int_{0}^{\infty} \frac{d t} {t^{\frac{8 - p}{2}}} \, e^{- \frac{\bar y^{2}}{2\pi \alpha' t}}  \,Z_{\rm (p + 1 - 2n) | (p + 1)} (t)\nn
&=& \Gamma_{\rm D(p + 1 - 2n)|D(p + 1)} (\bar y),
\eea
where in the last equality we have set $V_{1 + (p + 1) - 2 n} = V_{1 + p - 2 n} 2 \pi \bar a$ with $\bar a \to \infty$ and also $Z_{\rm (p + 1 - 2n)|(p + 1)} (t) = Z_{(p - 2 n)|p} (t)$.  This result is the expected one. In other words, we can apply this approach to the cylinder amplitude for the initial $D0|D(2n)$ system with a T duality once a time along one of the directions transverse to both the branes to obtain the general cylinder amplitude for a non-compactified ${\rm D(p - 2n)|Dp}$ system with $2n \le p \le 8$.  But this does not work in the reverse direction, i.e., starting from the cylinder amplitude for the ${\rm D(8 - 2n)|D8}$ system. 

\section{Conclusion and discussion}
In this paper, we compute the closed string tree cylinder amplitude between a D(p - 2 n) brane and a Dp brane  with $n = 0, 1, 2, 3$ and $2 n \le p \le 8$ for which there are $k \le p - 2 n$ longitudinal circle compactifications common to both branes with the respective radii $r_{\alpha_{1}}, r_{\alpha_{2}}, \cdots r_{\alpha_{k}}$ and $l \le 9 - p$ transverse circle compactifications with the respective radii $R_{j_{1}}, R_{j_{2}}, \cdots R_{j_{l}}$.  These two branes are placed parallel at a separation $\hat y^{\perp}$ along the non-compactified directions transverse to both branes and are also along each of the compactified transverse circles with a separation $|y^{j_{m}}|$. Note that there are Willson lines $y^{\alpha_{i}}$ turned on along the Dp brane worldvolume but there are no such on the D(p - 2n) brane for simplicity. 

With this amplitude, when the decompactification limits are taken, it gives the corresponding known non-compactified amplitude.  We explicitly check T duality along either a longitudinal or a transverse direction to both of these branes and the result meets the expectation. We explain the validity of the usual low energy T duality, which is usually employed in, for example,
in obtaining different supergravity D brane configurations from a given one, in obtaining the non-compactified cylinder amplitude when it is applied to a transverse direction to both of the branes while taking the compactified radius vanishing.  The rational behind this is that the low energy T duality does not take the momentum and winding modes into consideration and this can only be good if the compactification circle is taken along a transverse direction and the radius of this circle is taken to vanish as explained in the previous section.

 % If in two-column mode, this environment will change to single-columnich 
% format so that long equations can be displayed. Use
% sparingly.
%\begin{widetext}
% put long equation here
%\end{widetext}

\section*{Acknowledgments}
The author acknowledges the support by grants from the NNSF of China with Grant No: 12275264 and 12247103.

\end{document}